\documentclass[review]{elsarticle}

\journal{Computational Materials Science}

\usepackage{bigints}
\usepackage{graphicx}
\usepackage{subfig}
\usepackage{floatrow}
\usepackage{amsmath}

\usepackage{hyperref}
\hypersetup{
    colorlinks=true,
    linkcolor=blue,
    filecolor=magenta,      
    urlcolor=cyan,
}
\usepackage{epstopdf}
\begin{document}

\begin{frontmatter}

\title{Modeling of Radiation-Induced Defect Recovery in 3C-SiC Under High Field Bias Conditions}

\author[dep1]{Ricardo Peterson} \author[dep1,dep2]{Debbie Senesky}
\address[dep1]{Department of Electrical Engineering, Stanford University}
\address[dep2]{Department of Aeronautics and Astronautics, Stanford University\newline * Corresponding author e-mail address: rp3@stanford.edu}

\begin{abstract}
In this work, the implications of high field bias conditions in radiation-induced defect recovery in 3C-SiC crystals is studied. It is well known that transient heating effects (or thermal spikes) occur when energetic swift heavy ions (SHIs) deposit energy to the surrounding medium via ionization. Here, we explore the dynamics of this transient event under high background electric fields in 3C-SiC, which is what occurs when an ion strike coincides with field-sensitive volumes. In this study, we use the Ensemble Monte Carlo method to quantify how the energy deposition of the ionized regions change in response to high background electric fields. Subsequently, we study the relationship between the radiation-induced thermal spike and defect recovery using molecular dynamics simulations. We find that field strengths below the critical breakdown of wide bandgap devices are sufficient to exacerbate the localized heating, which subsequently enhances the defect recovery effect. This work is beneficial for 3C-SiC electronics and materials used in high radiation environments.
\end{abstract}


\end{frontmatter}

\section{Introduction}

The resurgence of space exploration has called for electronic devices and materials capable of withstanding extreme environments. One such environment is high particle radiation, which induces several measurable phenomena in semiconductor devices used in space-born systems. The first phenomenon is transient in nature, also known as a single event effect (SEE), which typically occurs when a radiation species penetrates sensitive regions of a device. The second is via material degradation due to long-term radiation exposure (e.g. total ionization dose), which slowly deteriorates the electrical behavior of the device. These two phenomena can ultimately lead to catastrophic mission failure due to device instability or unrecoverable damage. 

Silicon carbide, in comparison to silicon, has shown impressive operational stability under high temperature and radiation-rich environments\cite{NEUDECK}. But, in order to fully leverage the properties of SiC, predicting its behavior under such extreme conditions is required -- a feat that necessitates a profound understanding of a device's response to ion strikes. For example, it has been shown that a penetrating ion induces transient lattice heating (i.e., thermal spikes), which may exceed the melting/sublimation point of the host material \cite{TOULEMONDE2000903}. This especially occurs during swift heavy ion (SHI) radiation events, where inelastic interactions with the host material can locally produce ``hot electrons'' with temperatures on the order of 10$^4$ K. Empirical data and models have shown that such electron temperatures couple to the atomic system to produce exceedingly high lattice temperatures \cite{TOULEMONDE2000904}. This phenomenon alone is a cause for concern for operational stability of electronics, as it can produce single event transients that irreversibly modify the electrical properties \cite{LOPEZETAL}. 

For SiC in particular, annealing and recrystallization is a much more reported phenomenon for SHI exposure \cite{BenyagoubETAL, ZHANGETALL}. Multiple studies have demonstrated SHI-induced annealing in SiC for electronic energy depositions exceeding 10~keV/nm, and some as large as 33 keV/nm \cite{BenyagoubETAL, BACKMAN2013261}. Moreover, Zhang et al., reported a threshold value of 1.4 keV/nm, whereby defect annihilation in SiC was observed \cite{ZHANGETALL}. This low threshold value (which the authors termed the ``intermediate regime'' of ion types) is far more accessible to industrial accelerators and can be considered an avenue to develop low-temperature annealing techniques.

On the other hand, catastrophic failure was observed in SiC diode devices under high reverse biasing conditions \cite{Kuboyama1,Javanainen1,Abbate1}, where it was predicted that high field conditions amplified the thermal transient event, which cascaded into a permanently damaged device. Although the ion species and field conditions considered in these references far exceed the conditions considered in this work, it nonetheless stresses the need to explore this relationship further. Thus, we expand upon the implications of ionization events near high fields (i.e., sensitive volumes), where the average energy of ionized electrons is dramatically increased and consequently results in enhanced phonon emission. Although the defect annihilation from ionization-induced heating has been documented for SiC, there is limited work studying these events in the context of high electric fields. This is particularly important for low-Z number ions, which are far more abundant in radiation environments \cite{MEWALDT1994737} and more accessible to industrial accelerators.

In this study, we use molecular dynamics (MD) to simulate thermal spike events in SiC, which results in significant defect recovery. In particular, for the first time, we demonstrate how high background field strengths common in wide bandgap materials can augment the energy deposition of the radiation-induced thermal transient, and observe the degree to which defect recovery is enhanced. In the next section, we examine how the thermal spike profiles are calculated. In section 3, we discuss the Ensemble Monte Carlo (EMC) technique used to quantify the thermal energy density with varying electric fields. In section 4, we detail the MD technique used to determine its effects on defect recovery, and lastly we discuss our results in section 5 before the concluding remarks. 

\section{Thermal Spike Calculation}
For decades, spatial profiles of thermal transients induced by single event effects have been thoroughly investigated using the inelastic thermal spike model (ITSM) \cite{CHETTAH20092719}. We use this model herein, the results of which define the inputs of our EMC and MD methods that follow. The ITSM is based on coupled cylindrical heat transport equations, each representing the spatial and temporal evolution of the electron temperature, $T_e$, and atomic temperature, $T_a$. The heat transport equations are

\begin{align}
C_e(T_e) \frac{\partial T_e}{\partial t} &= \frac{1}{r}\frac{\partial}{\partial r} \left [rK_e(T_e) \frac{\partial T_e}{\partial r} \right] - g\cdot(T_e - T_a) + A(r,t)
\end{align}

and

\begin{align}
C_a(T_a) \frac{\partial T_a}{\partial t} &= \frac{1}{r}\frac{\partial}{\partial r} \left [rK_a(T_a) \frac{\partial T_a}{\partial r} \right] + g\cdot(T_e - T_a),
\end{align}

\noindent 
where the parameters $C$ and $K$ are the heat capacity and thermal conductivity, respectively, $r$ is the radial distance from the track center, and $t$ is time. The parameters $K_a$ and $C_a$ are a function of temperature and extracted from Ref.~\cite{SNEAD2007329}. The term $g \cdot (T_e - T_a)$ is the electron-phonon interaction term that defines the coupling between the two subsystems. The coupling constant $g$ was calculated from the relation $g = D_e C_e / \lambda^2$, where $\lambda$ is the electron mean free path and is deduced from the 3C-SiC band-gap ( $\sim 6$ nm) \cite{TOULEMONDE2000904}. The parameters $D_e$ and $C_e$ are the electron diffusivity and specific heat, respectively, and are taken to be constant ($C_e = 1$ J cm$^{-3}$ K$^{-1}$ and $D_e = 2$ cm$^2$ s$^{-1}$). This simplification is based on the premise that conductive electrons in insulators behave like electrons in metals \cite{TOULEMONDE2000904}. The initial energy distribution of the electrons is $A(r,t)$ is the source term (i.e. ion strike) defining the energy density deposited to the electron subsystem. We calculate this function using TOPAS (TOol for PArticle Simulation), a user-friendly software interface that is built on the GEANT4 Monte-Carlo toolkit~\cite{topas}, although analytical approaches that accurately approximate MC methods are often used \cite{TOULEMONDE2000904, WALIGORSKI1986309}.

\section{Field-dependent Hot Electron Energy Dissipation}

One main advantage of wide bandgap materials in electronic device design is their high breakdown electric field, which is on the order of $\sim 1$ MV/cm for 3C-SiC (and several fold higher for 4H-SiC and GaN). Although these properties are advantageous for high power device applications, the much higher electric fields enhance the sensitivity of the device and thus exacerbate any SEEs that may occur when operating in radiation environments \cite{Kuboyama1}. Specifically, in the event of an ion strike, the center of a cylindrical volume around the particle track becomes saturated with high-energy electrons, which subsequently relax by interacting inelastically with the lattice (i.e., in the form of phonons). In this work, our proposition is that this heating phenomenon is far more pronounced near field-sensitive volumes (e.g. Schottky interfaces at large biases, or channel-drain regions in field-effect transistors), where the full relaxation of energetic electrons is prolonged by the energy gain from an applied field. Although the ion strike alone can induce sufficient heating to alter the crystal structure of the host material, such events can be far more pronounced near field-sensitive volumes.

It is widely understood that ``hot electrons'' deposit most of their energy to the atomic subsystem via electron-optical phonon scattering. In order to quantify the emission of optical phonons as a function of background electric fields, the EMC method is employed, which uses a well-documented algorithm described elsewhere \cite{LUNDSTROM, VASILESKA}. For this method, we assume that the electric field is oriented along the $\langle 111 \rangle$ direction of the 3C-SiC crystal so that we ignore the redistribution of carriers among equivalent band-structure valleys (as in Ref. \cite{VASILESKA2}). This simplification allows us to use a single equivalent non-parabolic valley in our calculations, expressed by

\begin{align}
E(1+\alpha E) = \frac{\hbar^2 k^2}{2 m_c}
\end{align}

\noindent
where $\alpha = 0.323 \text{ eV}^{-1}$. The density-of-state mass $m_d = 0.346 m_0$ was used to determine scattering rates and $m_c = 0.313 m_0$ was used for the equations of motion. For our purposes, we focus on the total energy deposited (in the form of optical phonons) as a function of background electric field. The scattering mechanisms considered here are acoustic, polar-optical and two non-polar optical interactions (zeroth-order and first-order). Due to the simulation times of interest here ($<$ 5 ps) and high electron energies, the thermal energy deposition is defined by the emitted optical phonons (defined by electron-optical phonon scattering events), which is comparably much higher than the emitted acoustic phonons. Empirically, optical phonons couple to acoustic phonons, which initiates heat transport \cite{POPETAL}. However, this mechanism is ignored here since we are primarily interested in the initial conditions of the ion strike event and the subsequent energy deposited to the lattice (which is augmented by the electric~field).

The EMC simulation begins by initializing the energy distribution of 10,000 particles (representing electrons) such that the energy per unit length matches the electronic stopping power $S_e$, with a radial distribution approximating $A(r,t)$ from section 2. The total depth of the simulated particles was 500 nm. In the case of 50 MeV O+ in 3C-SiC, the net energy of the simulated particles is $1.60$ keV/nm.  During the EMC algorithm, a random number is drawn to determine the time of free flight for each particle. At the end of the free flight, a scattering operation may occur with a probability that is a function of the particle's energy. The particle's momentum after scattering (Eq. 3) is determined via the conservation of energy, $E'_{k} = E_k - \hbar \omega_{LO}$, which is the inelastic scattering process that defines the deposited thermal energy. Note that due to the stationary nature of an optical phonon, no dispersion is assumed. Thus, since heat transport is ignored and unnecessary during these time frames, the position of each particle (after scattering) is regarded as a quantum of the atomic thermal energy. The phonon energies have been set to $100$ meV and $120$ meV for nonpolar and polar-optical phonons, respectively \cite{NILSON}. The EMC simulation is run for 3 ps (well after steady-state is reached), after which the total energy of the emitted optical phonons is stored. Although phonon emission would continue past this simulation time (assuming high $E_z$), the energy density decreases substantially as the particles diffuse radially. The results are shown in Fig.~1 and Fig.~2. We observe in Fig. 2a that high energy carriers will continue emitting optical phonons at time scales several-fold higher than a typical energy relaxation time for hot electrons. By binning the system, we normalize using the 3C-SiC atomic density and plot the added thermal energy per atom (Fig. 2b). In other words, Fig. 2b plots the difference between the thermal energy per atom at non-zero and zero~$E_z$, which can be expressed as $\langle k_B T \rangle (E_z) - \langle k_B T \rangle (E_z = 0)$. This process is repeated for multiple electric fields for 50 MeV O+ and 50 MeV Al+, the data of which is used in the MD simulations that follow. The parameters used in our EMC simulation are presented in Table~1, and the ions considered are presented in Table~2.

 
\begin{table}[b!]
\caption{EMC modeling parameters \cite{NILSON}.} 
\centering 
\begin{tabular}{l c}
\hline\hline 
 Parameter & Value\\ [0.5ex] 
\hline 
Bandgap (eV) & 2.3\\
Nonparabolicity factor (eV$^{-1}$) & 0.323\\
Density (g/cm$^3$) & 3.21\\
Acoustic deformation potential (eV) & 16\\
Nonpolar optical phonon coupling:  & \\
\hspace{0.3cm} Zeroth-order (10$^8$ eV/cm) & 9\\
\hspace{0.3cm} First-order (eV) & 5.6\\
Nonpolar optical phonon energy (meV) & 100\\
Polar optical phonon energy (meV) & 120\\
Low-frequency dielectric constant & 9.7\\
High-frequency dielectric constant & 6.5\\
\hline 
\end{tabular}
\label{table:nonlin} 
\end{table}

\begin{table}[b!]
\caption{Ion specifications. The electronic stopping power S$_e$ is calculated using SRIM \cite{ZIEGLER20101818}.} 
\centering 
\begin{tabular}{l c c} 
\hline\hline 
 & O+ & Al+ \\ [0.5ex] 
\hline 
Ion Energy (MeV) & 50 & 50 \\
Ion Specific Energy (MeV/u) & 3.12 & 1.85\\
S$_e$ (keV/nm) & 1.60 & 4.0 \\
T$_{a,max}(r=0)$ (K) & 850 & 1400\\ [1ex] 
\hline 
\end{tabular}
\label{table:nonlin} 
\end{table}

\floatsetup[figure]{style=plain,subcapbesideposition=top}
\begin{figure}[h!]
  \hspace*{-2.4cm}\sidesubfloat[]{\includegraphics[clip,trim=.1cm 0cm .6cm 0cm,width=0.69\textwidth]{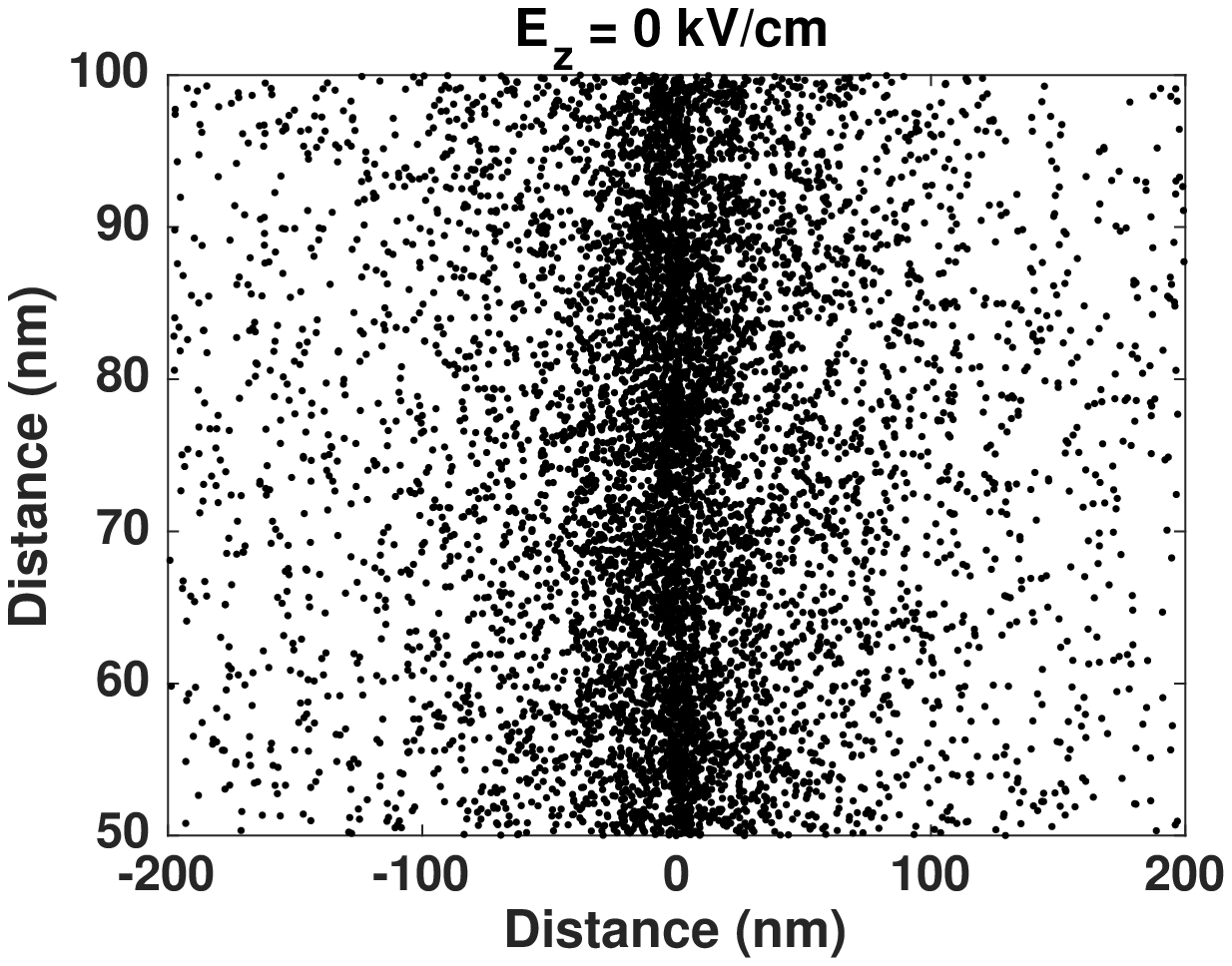}\label{fig:sub1}}~
  \hspace*{-.2cm}\sidesubfloat[]{\includegraphics[clip,trim=.7cm 0cm .9cm 0cm,width=0.64\textwidth]{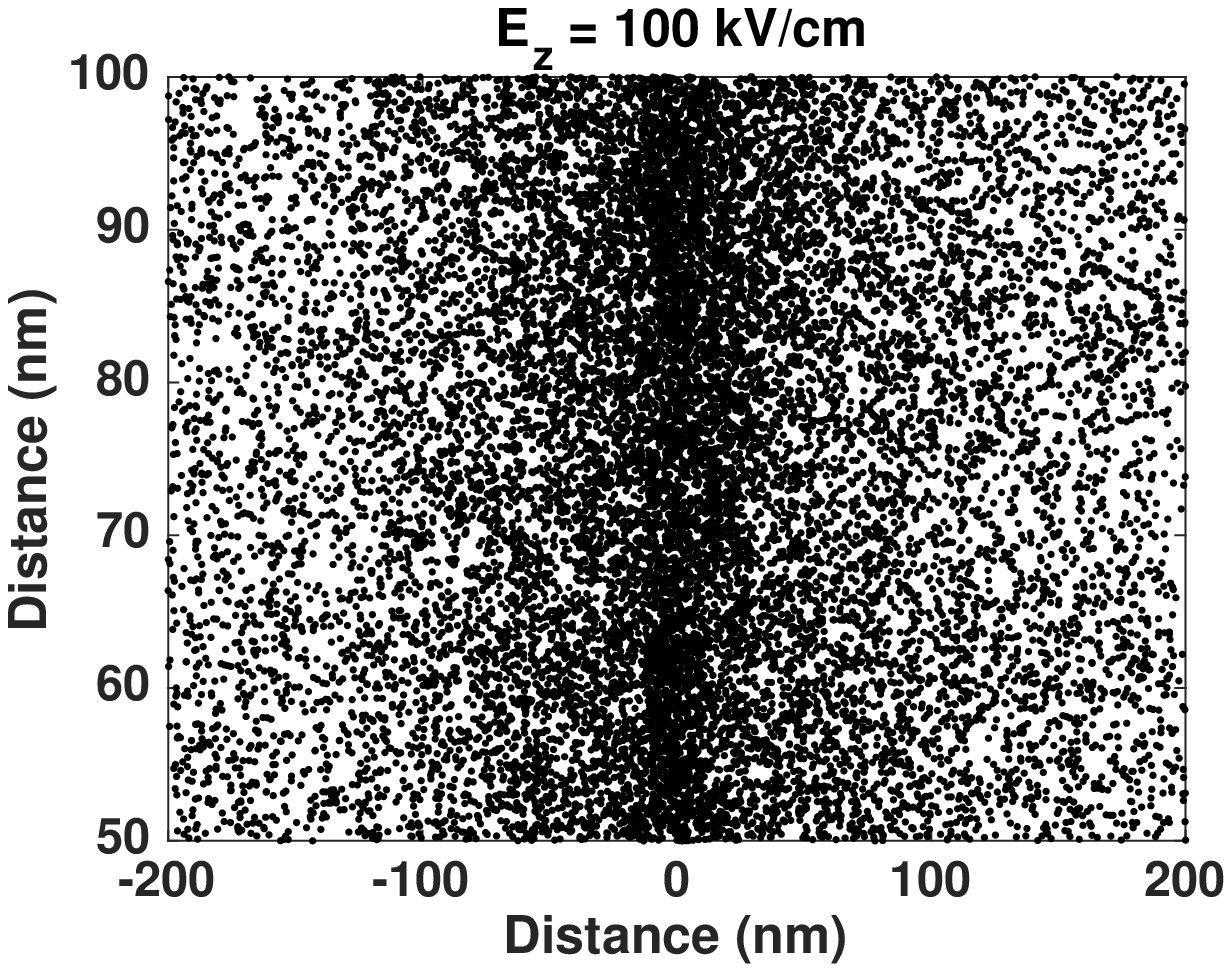}\label{fig:sub2}}%
  \caption{Illustration of optical phonon scattering positions in a 3C-SiC crystal as simulated by EMC. Images are snapshots at $t = 3$ ps. For contrast, only the phonons produced by 1\% of particles are shown.}\label{fig:test}
\end{figure}

\floatsetup[figure]{style=plain,subcapbesideposition=top}
\begin{figure}[h!]
  \hspace*{-2.6cm}\sidesubfloat[]{\includegraphics[clip,trim=0.1cm 0cm .2cm 0cm,width=0.70\textwidth]{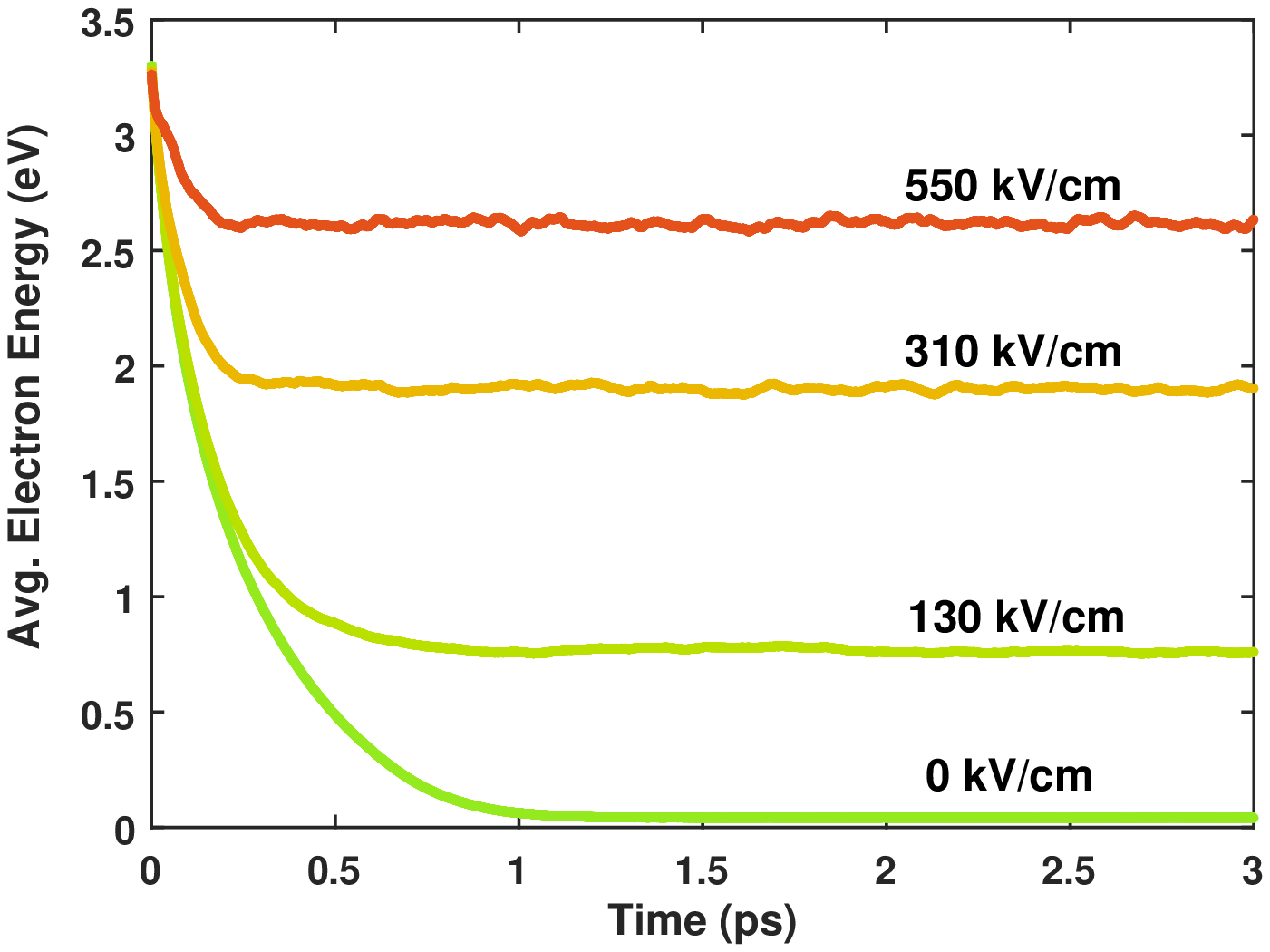}\label{fig:sub1}}~
  \hspace*{-.2cm}\sidesubfloat[]{\includegraphics[clip,trim=0cm 0cm 0cm 0cm,width=0.67\textwidth]{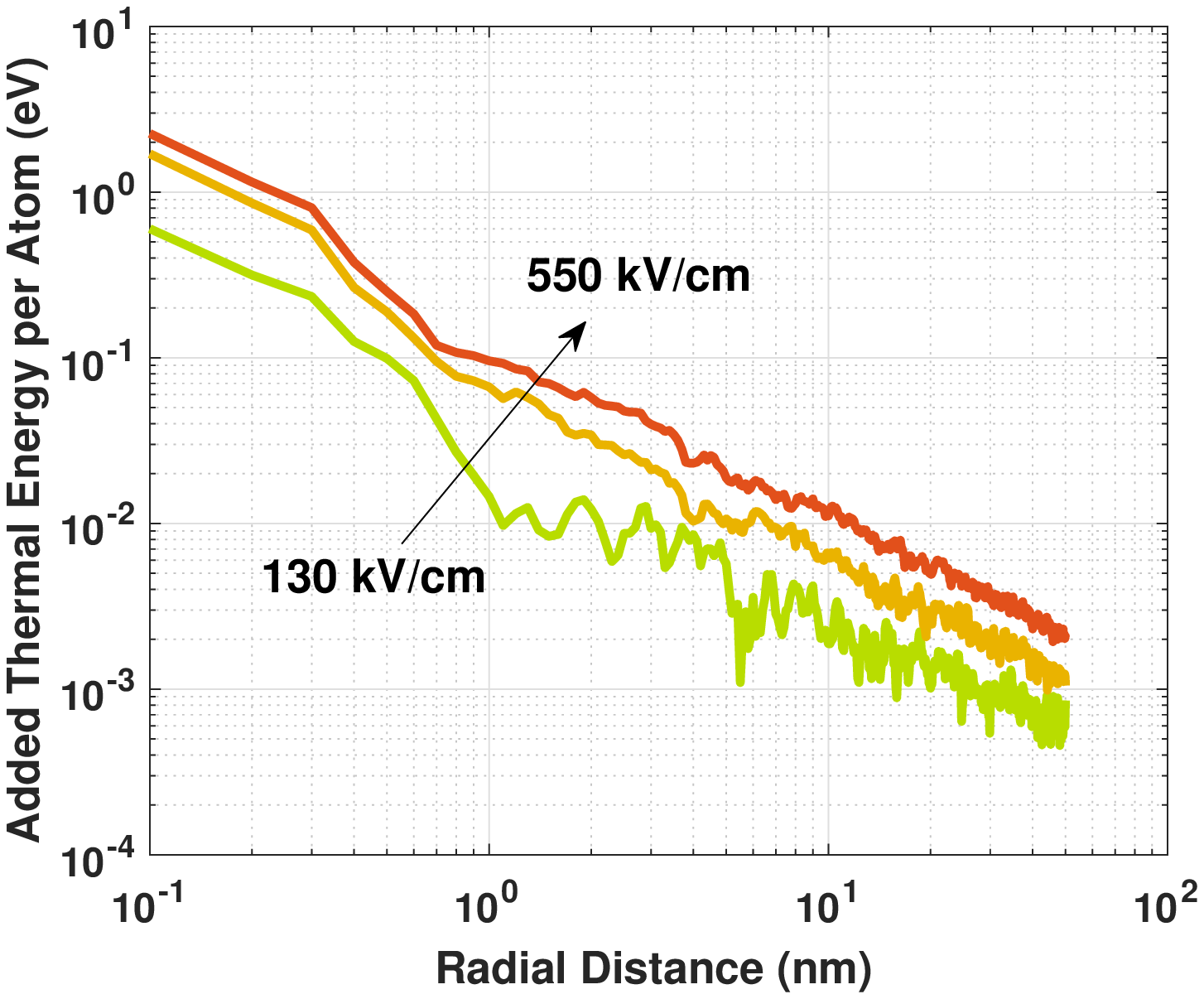}\label{fig:sub2}}%
  \caption{(a) Average electron energy at different electric fields, where $t=0$ corresponds to the moment of a 50 MeV O+ strike. (b) Added thermal energy per atom as a result of a non-zero electric field at $t=3$~ps (i.e. $\langle k_B T \rangle (E_z) - \langle k_B T \rangle (E_z = 0)$)}\label{fig:test}
\end{figure}

\newpage

\section{Molecular Dynamics Simulation}

The ITSM temperature distribution T$_{a,max} (r)$, which is induced by the ion strike (as calculated in section 2), is subsequently imported into our molecular dynamics simulations, which we employ using the Large-Scale Atomic/Molecular Massively Parallel Simulator (LAMMPS) \cite{PLIMPTON19951}. The function T$_{a,max}$ corresponds to the time when the thermal spike reaches its maximum temperature. The Si and C interactions are modeled by the hybrid Gao-Weber (GW)/ZBL potential \cite{gaoweber}, since it has previously been used to model defect annihilation \cite{ZHANGETALL,GAOWEBER3}, recrystallization \cite{DEBELLE}, and recovery due to swift heavy ions \cite{BACKMAN2013261}. Hence, this potential is expected to accurately describe the damage production and recovery processes. The simulated 3C-SiC system was 36 $\times$ 36 $\times$ 6 nm$^3$ containing 768,000 atoms. The system was defined to keep the number of atoms and volume constant while controlling the temperature through a 2 nm-thick Berendsen thermostat in the x-y boundaries, which acts as a thermal reservoir. A fixed timestep of 0.2~fs was used.

Prior to importing the thermal spike, the system energy was minimized using the energy minimization function in LAMMPS. Then, defects were created by initializing the 3C-SiC system with thousands of randomly generated Frenkel pairs (FPs). After the creation of FPs, the system energy was once again minimized as the generated defects migrated to metastable positions. This process was repeated until approximately 2\% of the system was defective (by modifying the initial set of displaced atoms). The defects were defined using a built-in Voronoi method in LAMMPS, where a polyhedral volume is defined around each atom in the pristine 3C-SiC. Thus, one atom per Voronoi volume implies a pristine crystal. The simulator identifies a vacancy if no atom is present in the volume, whereas an interstitial is identified if 2 or more atoms are present. Incidentally, the thermal dissipation properties of the lattice is expected to have a dependence of the lattice disorder. However, we ignore this dependence here since it has been demonstrated to have little effect on the defect recovery behavior \cite{BACKMAN2013261}. 

To simulate the thermal spike event, the system was binned into multiple cylindrical volumes so that the atoms in each bin could be separately operated on. The average kinetic energy of the atoms in each bin was defined such that the MD temperature profile best fitted T$_{a,max} (r)$ from the ITSM, as shown in Fig.~3 for the case of 50 MeV O+. The timestep $t = 0.8$~ps was selected for Fig.~3 since the temperature profile is approximately at its maximum. To represent an increase in $E_z$, the EMC method was used to extract the added thermal energy per atom (Fig.~2b). This profile was subsequently used at the initialization of the MD simulation, where the kinetic energy of atoms inside the cylindrical bins was increased in accordance to the calculations from EMC. This thermal spike simulation was run for a total of 50~ps, after which the defect concentration was characterized (extending the simulation time for longer periods did not noticeably affect the lattice disorder). The result is plotted in Fig.~4 at $t = 0.8$~ps for up to $1000$ kV/cm.

Although the GW potential has been demonstrated multiple times to model defect dynamics in MD, it is critical to question how our results relate to the activation energies associated with defect production and recovery. One study has demonstrated that close-range FP recombination barriers in GW can range between 0.22 eV and 1.6 eV \cite{GAOWEBER3}, which was regarded as relatively low according to \textit{ab initio} results from reference \cite{LucasETAL}. More generally, however, computational and experimental values obtained for recombination barriers for both silicon and carbon FPs vary, which is likely caused by inconsistent methods used in \textit{ab initio} calculations \cite{MZhengETAL}. For this reason, we believe it is appropriate to simulate a thermal spike event using the environment-dependent interatomic potential (EDIP), which exhibits activation energies higher than those of GW, while accurately describing bulk properties and point defects of 3C-SiC \cite{LucasETAL2}. However, a detailed comparison between these two MD potentials is beyond the scope of this paper. 

\floatsetup[figure]{style=plain,subcapbesideposition=top}
\begin{figure}[t!]
  \hspace*{-2.6cm}\sidesubfloat[]{\includegraphics[clip,trim=0.2cm 0cm .2cm 0cm,width=0.63\textwidth]{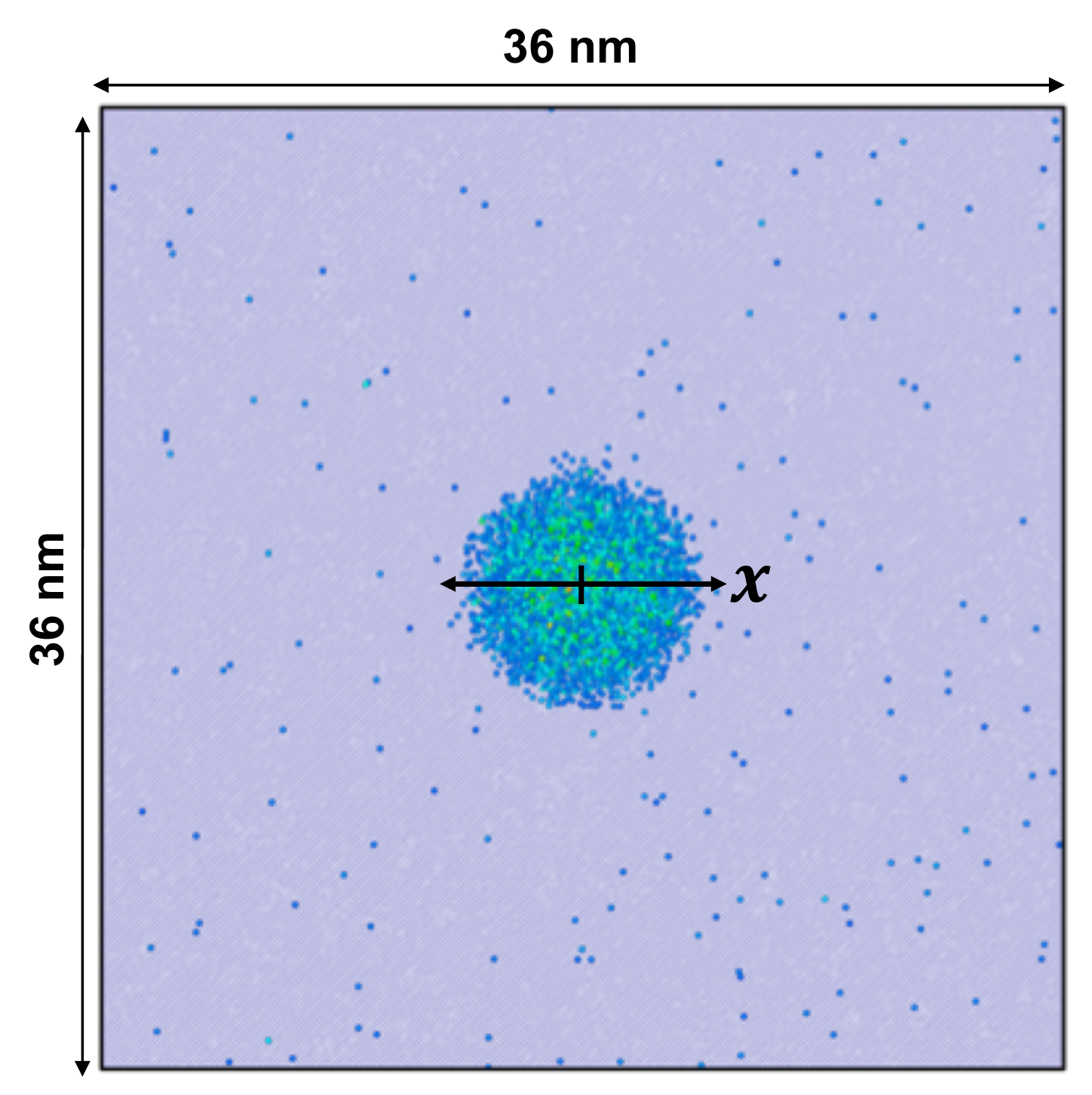}\label{fig:sub1}}~
  \hspace*{0cm}\sidesubfloat[]{\includegraphics[clip,trim=0.2cm 0cm 0cm 0cm,width=0.63\textwidth]{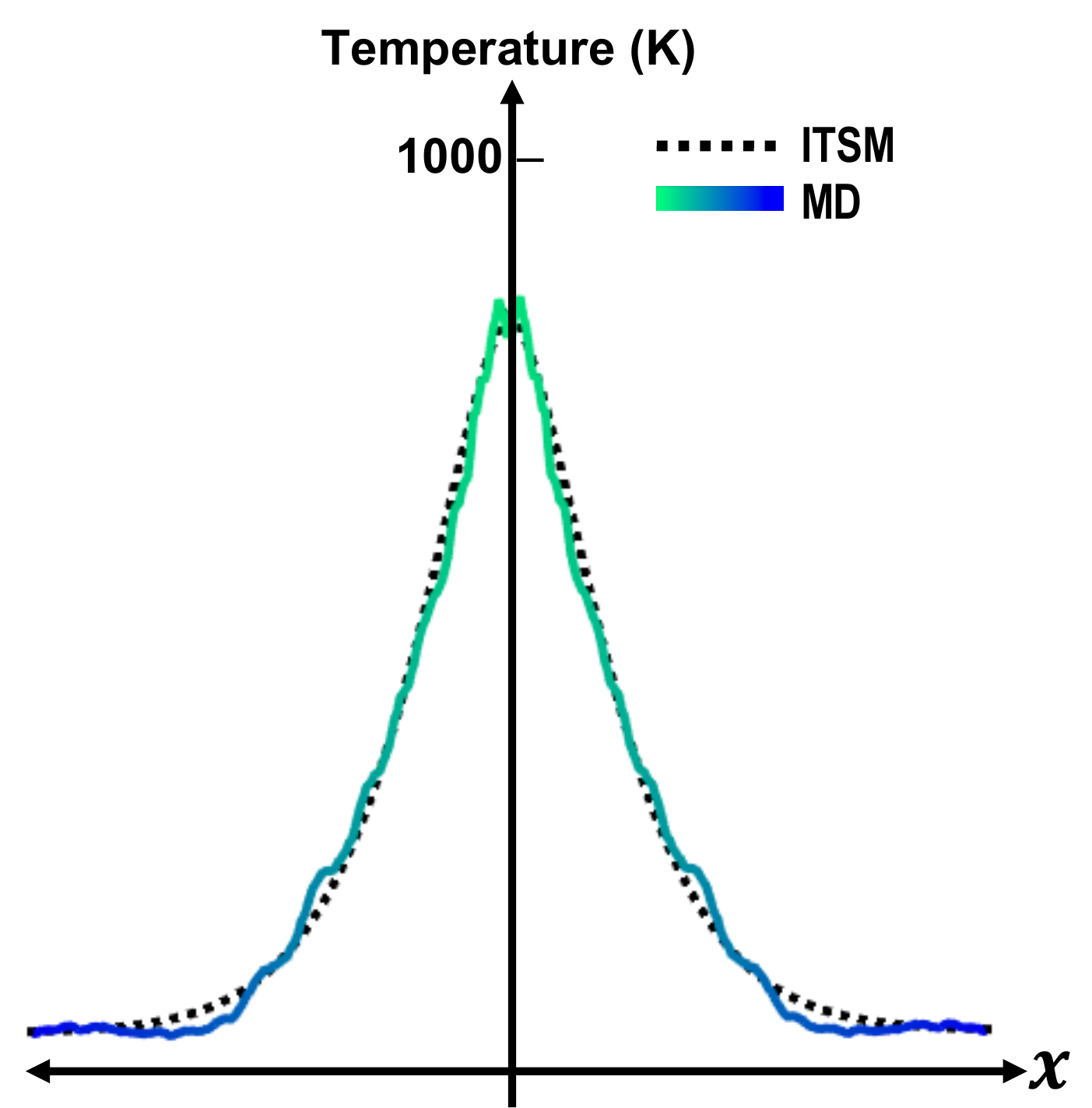}\label{fig:sub2}}%
  \caption{MD thermal spike spatial visualization of 50 MeV O+ ion impact in a 3C-SiC crystal for zero electric field (hence the profile matches the ITSM exactly). The MD temperature profile is taken at $t = 0.8$~ps. The lighter-colored atoms correspond to higher kinetic energies. In (a) atoms at room temperature $k_B T$ are made semi-transparent. The center region illustrates a thermal spike event, with a temperature profile shown in (b). }\label{fig:test}
\end{figure}

\floatsetup[figure]{style=plain,subcapbesideposition=top}
\begin{figure}[t!]
  \hspace*{-2.4cm}\sidesubfloat[]{\includegraphics[clip,trim=0cm 0cm 0cm 0cm,width=0.72\textwidth]{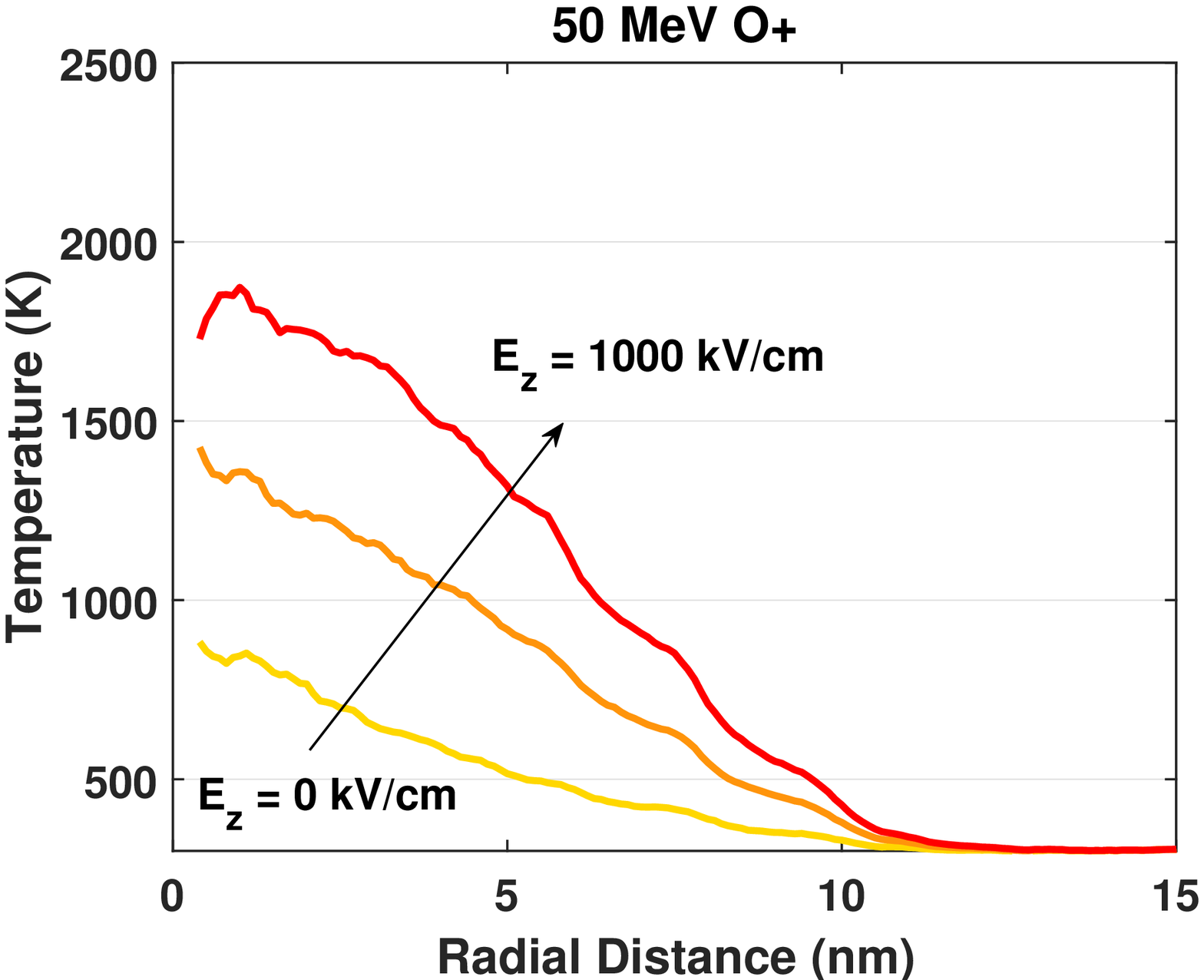}\label{fig:sub1}}~
  \hspace*{-0.7cm}\sidesubfloat[]{\includegraphics[clip,trim=2.6cm 0cm 0cm 0cm,width=0.63\textwidth]{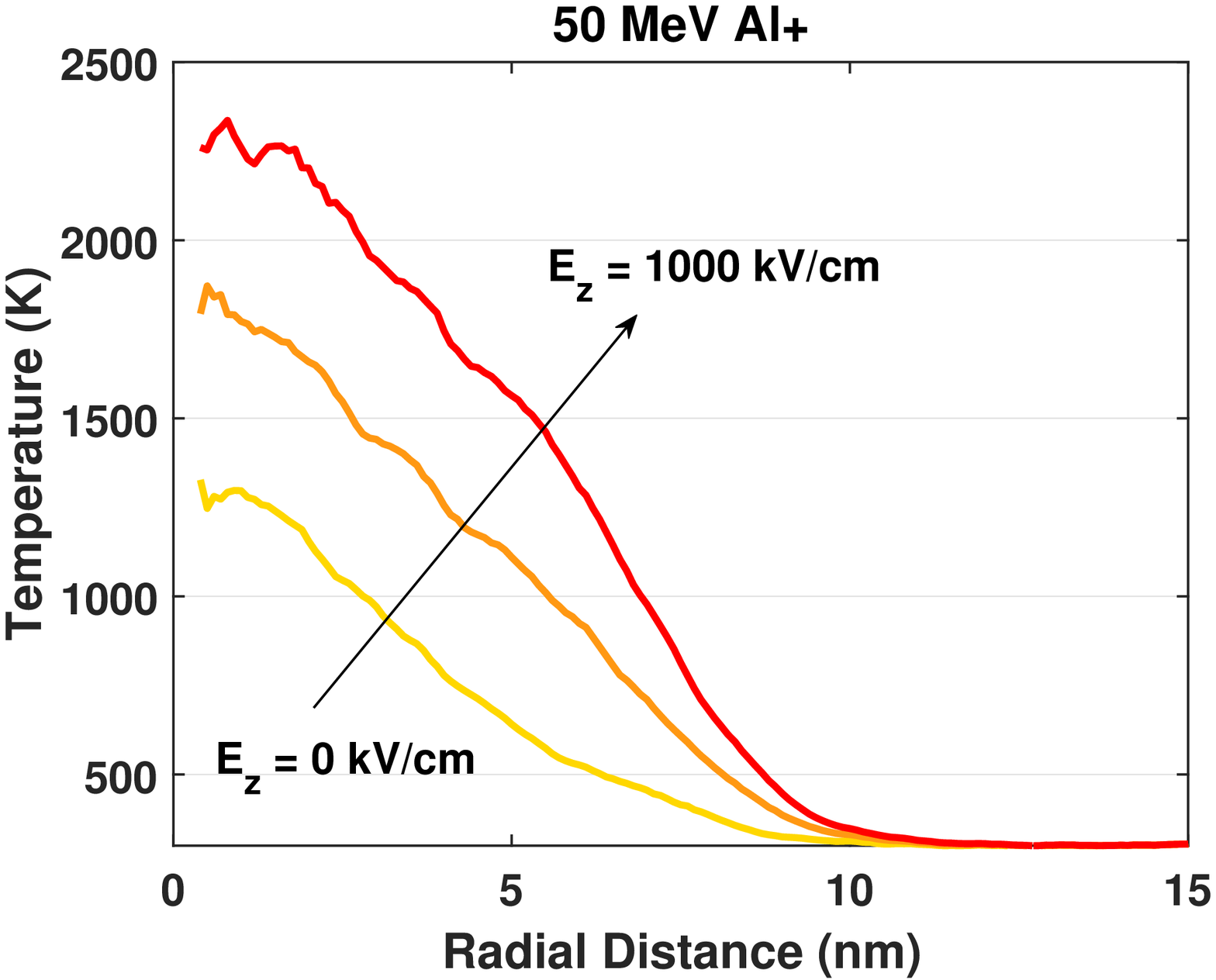}\label{fig:sub2}}%
  \caption{Thermal spike output of MD at $t = 0.8$ ps for 3C-SiC crystal.}\label{fig:test}
\end{figure}

\section{Results and Discussion}
	Two different ion species (50 MeV O+ and 50 MeV Al+) were considered in this work, which were selected according to their relatively larger abundance and accessibility in radiation environments \cite{MEWALDT1994737, Badhwar}. Analysis of the lattice disorder reveals significant defect recovery induced by the thermal spike events. Fig.~5 shows the lattice disorder of the relaxed system after each event, radially averaged around the center of the ion track. The relative disorder is calculated by normalizing the number of defects over a given volume by the initial defect concentration. Hence, the mean value of the radial distribution ``Before Strike" is equal to unity. In the case of zero electric field, lattice healing is observed for both ion species. In particular, the slight recovery observed for O+ ($1.60$~keV/nm) is consistent with the data reported in \cite{ZHANGETALL}, where a threshold value for recovery was found to be $1.40$~keV/nm. It is worth noting that the nature of defect recovery is sensitive to the initial state of the defective lattice \cite{SAHOO201845}. Thus, depending on the initial concentration of Si and C defects, the recovery process may differ significantly.

    More importantly, Fig. 5 and Fig. 6 show significant defect recovery as $E_z$ increases. For instance, in the case of O+, $E_z \sim 1000$ kV/cm decreases the final disorder by 33\%, as shown in Fig. 6a. This field is approximately at the breakdown level of 3C-SiC, and several MV/cm below the breakdown field of other SiC polytypes. The lattice disorder continues to decrease with $E_z$, although the silicon defect recovery appears to converge (Fig. 6b). Indeed, the concentration of carbon defects are known to be significantly greater than silicon, which is explained in part by the lower vacancy/interstitial energy barriers \cite{GAOWEBER3, SAMOLYUK201583}. This explains the difference in the fractional contribution of disorder for each element, as shown in Fig. 6b. 
    
    However, a thermal spike simulation using the EDIP potential yields a significant difference in the defect recovery effect, as also shown in Fig.~6a. For an identical thermal spike profile, negligible defect recovery is observed after a 50~MeV O+ strike when using EDIP.  However, when presuming $E_z \sim 1000$~kV/cm, the defect recovery effect is observed, although the magnitude is about 20\% lower when compared to the Gao-Weber potential. This difference is likely attributed to the higher activation energies associated with the EDIP potential, as reported in \cite{LucasETAL2}. We stress, however, that a dedicated study on the differences between these potentials for this application is needed and encouraged (e.g. \cite{SAMOLYUK201583}).
    
\floatsetup[figure]{style=plain,subcapbesideposition=top}
\begin{figure}[b!]
  \hspace*{-2.6cm}\sidesubfloat[]{\includegraphics[clip,trim=0cm 0cm 0cm 0cm,width=0.66\textwidth]{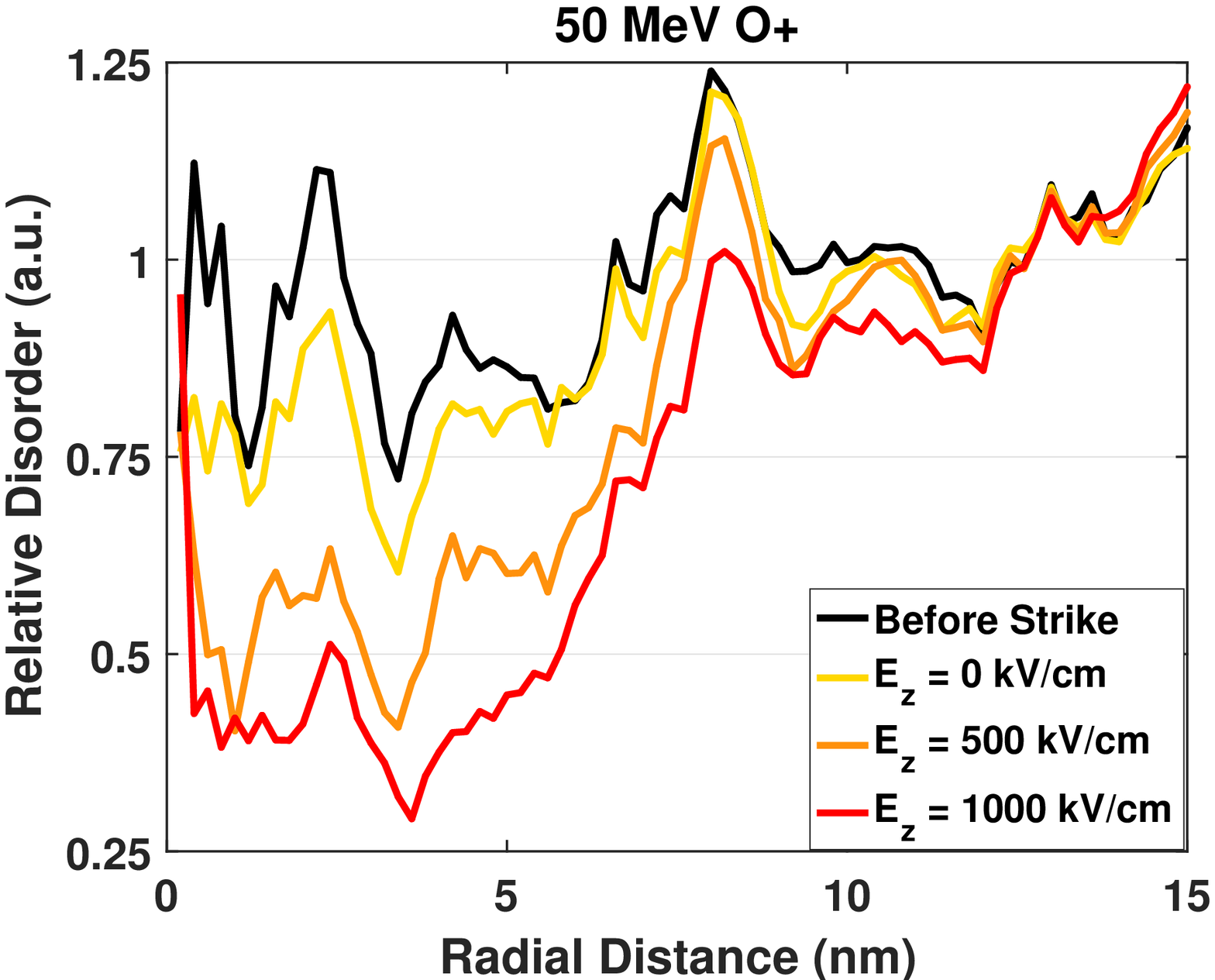}\label{fig:sub1}}~
  \hspace*{-0.8cm}\sidesubfloat[]{\includegraphics[clip,trim=0.88cm 0cm 0cm 0cm,width=0.64\textwidth]{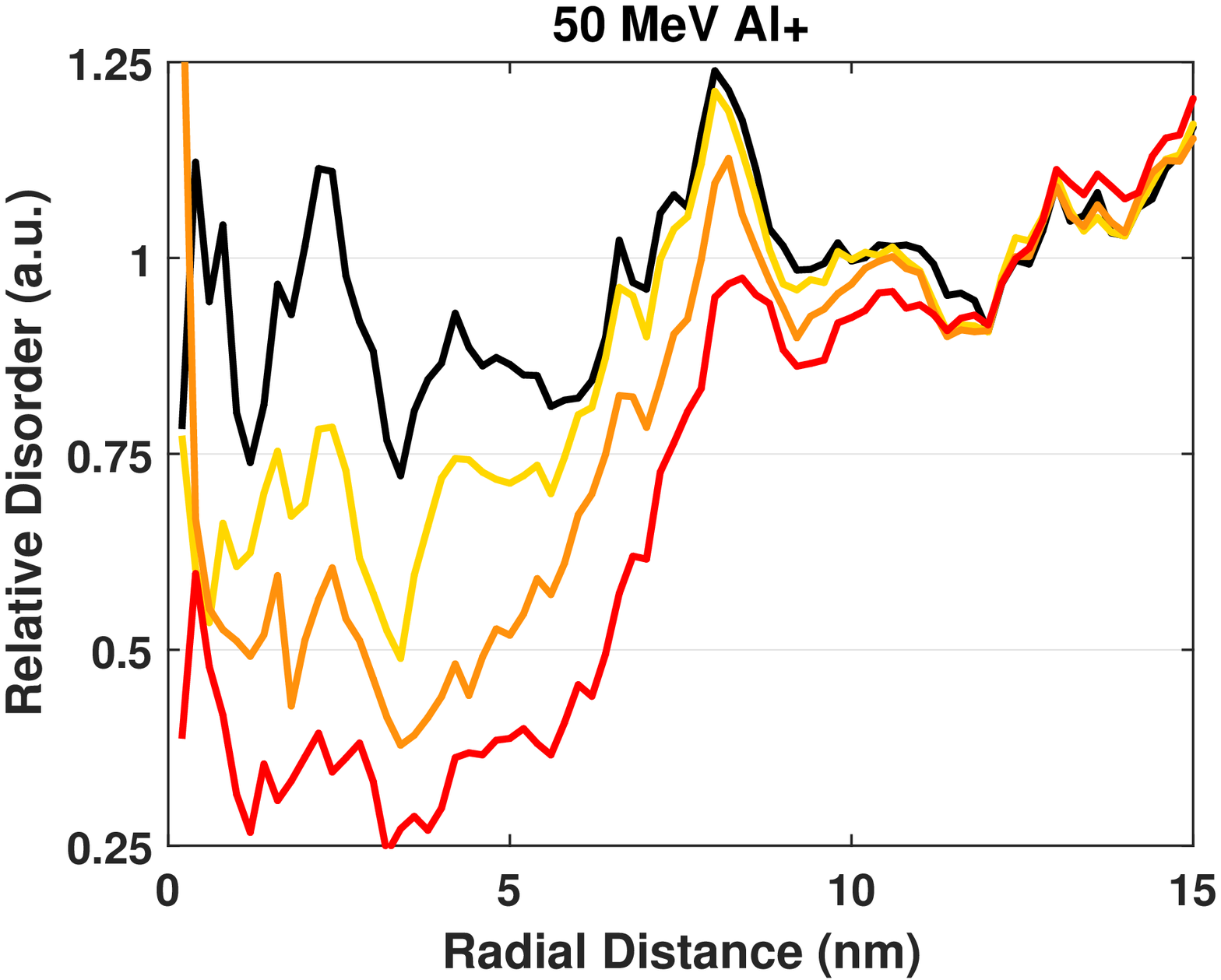}\label{fig:sub2}}%
  \caption{Relative disorder of SiC system at thermal equilibrium ($t = 50$~ps). The black line represents the disorder of the defective lattice before the ion strike. The colored lines represent the disorder of the lattice at increasing $E_z$.}\label{fig:test}
\end{figure}
    
    It is worth noting that the initial charge distribution as defined by $A(r,t)$ in section 2 is a close reflection of rate of Coulomb scattering. For example, the electrons at the center of the ion track 
    (where $n_e\sim$ 10$^{23}$ cm$^{-3}$) exhibit greater Coulomb scattering than the electrons in the periphery of the ion track (where $n_e\ll$~10$^{21}$~cm$^{-3}$). Each elastic collision inhibits the energy that electrons could otherwise gain ballistically. Thus, it follows that drifting carriers gain more energy at the peripheries of the excited region. This can be consequential for ion strikes that deposit energy over larger volumes (as evidenced by the velocity effect \cite{Meftah_etal}), where Coulomb scattering can be comparably much less. 
    
\floatsetup[figure]{style=plain,subcapbesideposition=top}
\begin{figure}[b!]
  \hspace*{-2.3cm}\sidesubfloat[]{\includegraphics[clip,trim=0cm 0cm 0cm 0cm,width=0.67\textwidth]{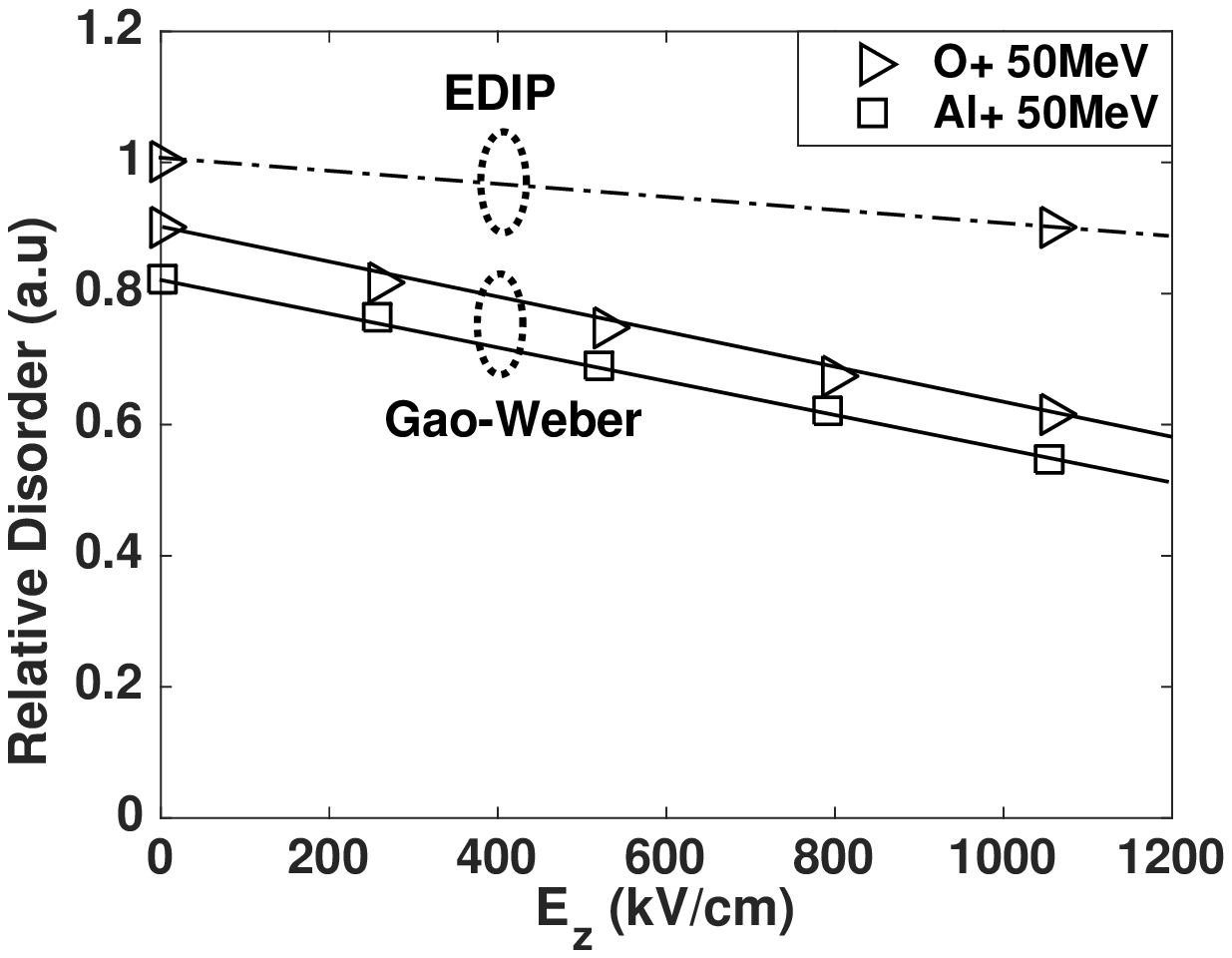}\label{fig:sub1}}~
  \hspace*{-0.8cm}\sidesubfloat[]{\includegraphics[clip,trim=0.8cm 0cm 0cm 0cm,width=0.618\textwidth]{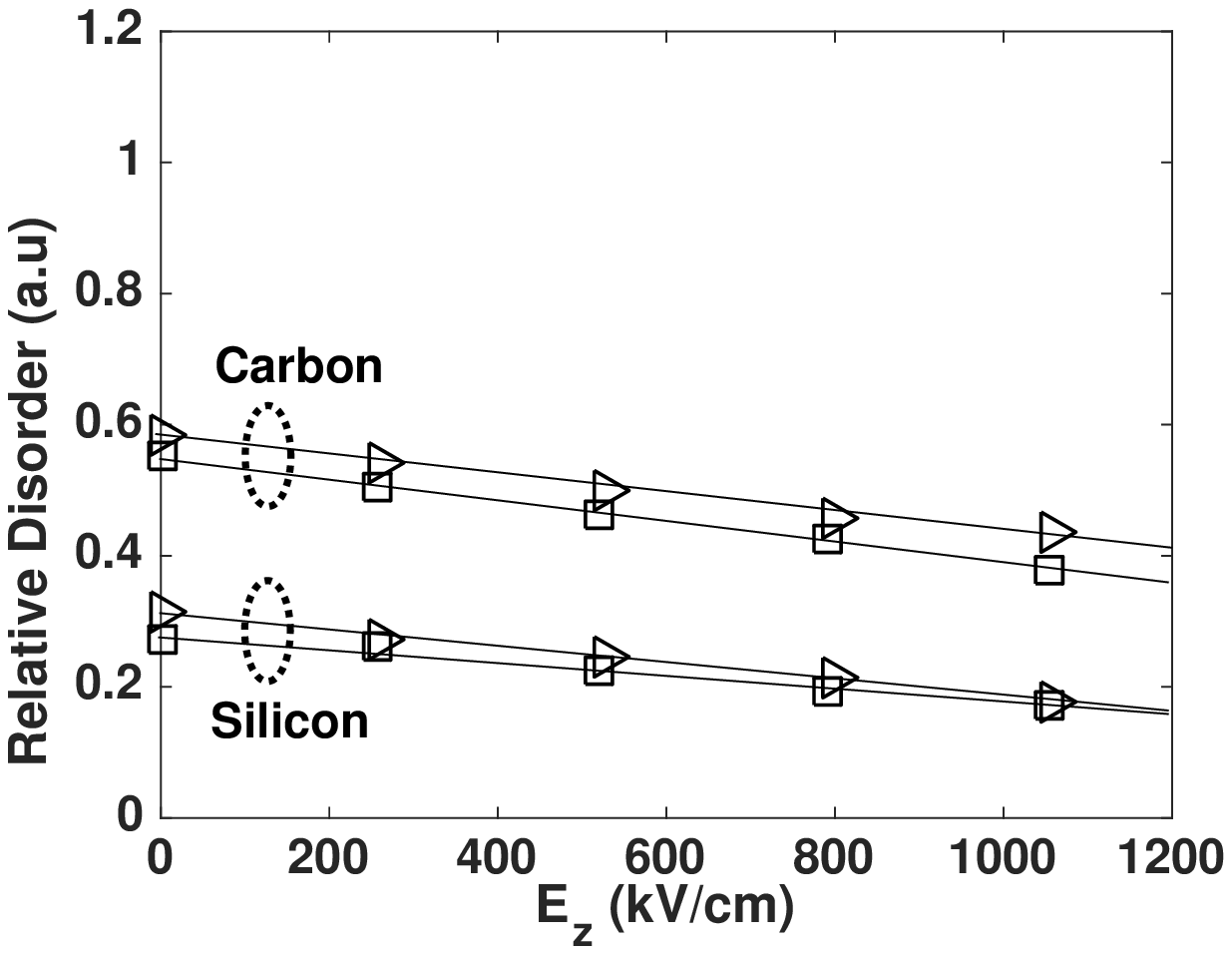}\label{fig:sub2}}%
  \caption{Relative disorder of 3C-SiC lattice at thermal equilibrium after thermal spike at increasing $E_z$. (a) A $\sim$ 33\% decrease in lattice disorder is observed as $E_z$ increases to $\sim$~1000~kV/cm. The EDIP potential \cite{LucasETAL2}, on the other hand, exhibits a $\sim$ 10\% decrease over the same field range. (b) The contributions of disorder for carbon and silicon. The disorder was averaged over a 10 nm radius cylindrical volume around the ion path.}\label{fig:test}
\end{figure}
    
    These results show us that lighter ions ($S_e < 5$ keV/nm) are sufficient to induce defect recovery up to about $\sim 8$ nm from the track center. This translates to a modest ion fluence on the order of $10^{12}$ cm$^{-2}$ if one wishes to anneal large sample regions. Previous work has demonstrated defect recovery in 3C-SiC, which include annealing induced by heavy and highly energetic ions (e.g., Pb ions of $33$~keV/nm \cite{BACKMAN2013261}) to much lighter and less energetic ions (e.g., $1.4$ keV/nm~\cite{ZHANGETALL}). Here we reiterate this phenomenon while demonstrating that, by increasing the average electron energy via an applied field bias, this recovery effect can be enhanced. 
    

\section{Conclusion}
	Localized heating as a consequence of radiation-induced ionization has been studied for decades, with implications that vary widely across different host materials. For 3C-SiC in particular, defect recovery resulting from heavy ion irradiation has consistently been reported. Here we report that this phenomenon can amplify in the presence of high background fields, which is a direct consequence of the high steady-state electron energy. We also hypothesize that this higher field response is more pronounced in regions with lower carrier densities, such as in ionized regions from high-velocity ions (which deposit energy over larger volumes). This work can be insightful for SiC-based applications, such as for fuel coating and structural components in nuclear reactors, or for SiC-based electronics operating in radiation-rich environments. Future work is needed to experimentally demonstrate the simulated results of this work. 
	
\section*{Acknowledgements}
	This material is based upon work supported by the National Science Foundation Graduate Research Fellowship under Grant DGE-1656518. Any opinions, findings, and conclusions or recommendations expressed in this material are those of the authors and do not necessarily reflect the views of the National Science Foundation. Some of the calculations in this work was performed on the Sherlock cluster. We would like to thank Stanford University and the Stanford Research Computing Center for providing computational resources and support that contributed to these research results.

\section*{Data Availability}
The raw and processed data required to reproduce these findings are available to download from \href{http://dx.doi.org/10.17632/wj2cy3kvy4.1}{doi:10.17632/wj2cy3kvy4.1}.

\newpage

\section{References}
\bibliography{referencesFile}

\end{document}